\documentclass[prx, twocolumn, showpacs, preprintnumbers, amsmath, amssymb, superscriptaddress, aps,longbibliography]{revtex4}
\usepackage{amssymb}
\usepackage{latexsym}
\usepackage{graphicx}
\usepackage{hyperref}
\usepackage[english]{babel}
\hypersetup{linktocpage,,colorlinks=true}
\usepackage{color}
\usepackage{epsfig}

\usepackage{dcolumn}
\usepackage{amsmath}
\usepackage{amsfonts}
\usepackage{bm}

\newcommand{\be}{\begin{equation}}
\newcommand{\ee}{\end{equation}}
\newcommand{\bea}{\begin{eqnarray}}
\newcommand{\eea}{\end{eqnarray}}

\newcommand{\p}{\partial}
\newcommand{\s}{\sigma}

\newcommand{\la}{\langle}
\newcommand{\ra}{\rangle}
\newcommand{\rd}{\mbox{d}}
\newcommand{\ri}{\mbox{i}}
\newcommand{\re}{\mbox{e}}

\renewcommand{\vec}[1]{{\bm #1}}

\begin{document}
\title{A Tractable Model of Pair Density Wave}
\author{ A. M. Tsvelik}
\affiliation{Division of Condensed Matter Physics and Materials Science, Brookhaven National Laboratory, Upton, NY 11973-5000, USA}
 \date{\today } 
 
 \begin{abstract} 
 The letter describes a simple mechanism for superconducting pairing with finite wave vector (Pair Density Wave) based on weak coupling. The mechanism  is illustrated with a quasi one-dimensional model relevant to certain class of quasi one-dimensional materials. Within this model pair and charge density wave order parameters are intertwined emerging as components of the same matrix field and are related by particle-hole symmetry. 
  \end{abstract}


\maketitle
  The discovery of exotic two-dimensional superconductivity in La$_{1.875}$Ba$_{0.125}$CuO$_4$ \cite{Tranquada} later attributed to condensation  of superconducting pairs with finite momentum \cite{Kivelson}, has triggered attention to the phenomenon of Pair Density Wave (PDW). Contrary to earlier suggestions by Fulde and Ferell \cite{FF} and Ovchinnikov and Larkin \cite{OL} the observed PDW was formed  in zero applied field and without any time reversal symmetry breaking. Since then zero field  PDWs  have been observed in many materials (see \cite{review} for review and more recently \cite{UTe2}). 
  
   Although the formation  PDW formation in LBCO is likely to be triggered by strong interactions, I would like to argue that contrary to the general belief expressed in  \cite{review}, the requirement of strong interaction is not that strict.  This argument  constitutes the main point of the current paper. Indeed, the singularity in the particle-particle channel may emerge for Fermi liquid with multi-sheet Fermi surface provided the nesting condition is fulfilled: 
  \bea
  \epsilon_a({\bf k}) = -\epsilon_b({\bf k} + {\bf Q}), \label{PH}
  \eea
  which  also leads to a singularity in the particle-hole channel. If these conditions are fulfilled the PDW can develop at weak coupling and it is likely either to compete or to be intertwined with Charge Density Wave (CDW) such that they 
  can be viewed as components of the same matrix order parameter 
 \bea
 \hat g = \left(
 \begin{array}{cc}
 \Delta_{cdw} & \Delta_{pdw}\\
 \Delta^*_{pdw} & \Delta^*_{cdw}
 \end{array}
 \right).\label{g}
 \eea
This is not to be confused with such pairing which exists in such materials as iron-based superconductors which also have Fermi surfaces with multiple sheets. Though the interactions between the sheets are believed to play a decisive role in the formation of the superconductivity, the pairs are arranged from  quasiparticles from the same sheet and have zero momentum (see, for example, \cite{chubukov}). 
  
  Since in the presence of inversion symmetry  pairing at zero momenta still remains a possibility, the realization of the above  weak coupling scenario requires  the inter-sheet pairing interaction to dominate over the one at zero momentum.  There is a positive evidence that both this condition and the nesting ones (\ref{PH})  can be realistically fulfilled. It comes  from the experiments  on  a quasi one-dimensional materials - molybdenum blue bronzes A$_{0.3}$MoO$_3$  where the chemical potential is crossed by two slightly warped bands with 3/4-filling \cite{band1,band2}. The ARPES measurements  on the blue bronze compound K$_{0.3}$MoO$_3$ \cite{blue} demonstrate both the nesting (\ref{PH}) between different Fermi surfaces and the dominance of the inter-sheet interaction leading to the formation of incommensurate CDW (see also \cite{2007}). These findings serve as a proof of principle. The subsequent analysis  suggests that PDW state in the blue bronze may lurk very close.   It is therefore concievable  that there are similar materials  with slightly different interactions which may form PDWs.  Being inspired by this example  I reconsider the model used to describe the blue bronze \cite{blue} and demonstrate that it describes a formation of  intertwined PDW and CDW orders  at weak coupling.   
 
{\it Model}. 
 We will consider a quasi-1D model describing chains with weak interchain tunneling. Each chain contains electrons with FS's with Fermi vectors $k_{F,a}$ ($a=1,2$).  All  interactions are supposed to be  sufficiently weak so that one can  linearize the quasiparticle spectrum at the corresponding FS. Then the  Hamiltonian density for each chain can be expressed in terms of right- ($R^+,R$) and left ($L^+,L$) fermion operators: 
\bea
&& {\cal H} = \ri v_a \left( -R^+_{a\s}\p_x R_{a\s} + L^+_{a\s}\p_x L_{a\s}\right) + \nonumber\\
&& \gamma_{ab}\rho_a\bar\rho_b + g_{ab} {\bf J}_a\bar{\bf J}_b, \label{model}\\
&& {\bf J}_a = R^+_{a\s}\vec\s_{\s\s'}R_{a\s'}, ~~\bar{ \bf J}_a = L^+_{a\s}\vec\s_{\s\s'}L_{a\s'}, \nonumber\\
&& \rho_a = R^+_{a\s}R_{a\s}, ~~ \bar\rho_a = L^+_{a\s}L_{a\s}.
\eea
We  consider the case $g_{ba} = g_{ab} > 0$ when the corresponding  current-current interactions are marginally relevant. 

  The simplest way to study model (\ref{model}) is by bosonization technique. Since the most of calculations are standard, I relegate most of them to Appendix.  The analysis is greatly simplified by the fact that the spin and charge sectors of the model are independent and Hamiltonian (\ref{model}) splits into two commuting parts: 
  \bea
  H = H_{charge} + H_{spin}.
  \eea
  The interaction of the spin currents generate  gaps in the spin sector, the charge one remains gapless and can be represented as a sum of four chiral Gaussian models describing dynamics of left- and right moving fields $\varphi_{a,c}$  and $\bar\varphi_{a,c}$ ($a=1,2$) which compactification radii are determined by couplings $\gamma_{ab}$. Depending on whether max$\{g_{11},g_{22}\}$ is smaller or greater than $g_{12}$ the order parameters are either off-diagonal in the sheet indices or diagonal \cite{phase}. The former case corresponds to PDW formation. There are two sets of OPs with different parity, corresponding (in terms of classification of \cite{review}) to unidirectional PDW:
\bea
&& \Delta_{pdw}(+,Q_-, ) = \ri R_{1}\s^yL_{2}, ~~Q_-= k_{F1}- k_{F2}, \nonumber\\
&& \Delta_{pdw}(-, -Q_-) = \ri R_{2}\s^yL_{1},\\
&& \Delta_{cdw}(+,Q_+) = R^+_1L_2,  ~~Q_+ = k_{F1}+k_{F2}\nonumber\\
&& \Delta_{cdw}(-,Q_+) = R^+_{2}L_1
\eea
If $g_{12} >> $max$(g_{11}, g_{22})$ the latter interactions can be neglected and the spin sector gets split into two parts with opposite chirality: 
$
H_{spin} = H^{(+)} + H^{(-)}$. The model then admits exact solution \cite{Andrei} and many of its correlation functions can be calculated (see \cite{tsvelik}, \cite{blue}) where for the simplest case of two coinciding Fermi velocities $v$ the corresponding Hamiltonians are given by the well familiar sine-Gordon models (see Appendix A):
\bea
&& H^{(\pm)} = \int \rd x \Big\{(1 + g/2\pi)[v^{-1}(\p_x\Theta_s)^2 + v(\p_x\Phi_s)^2] + \nonumber\\
&& \frac{g}{(2\pi a_0)^2}\cos(\sqrt{8\pi}\Phi_s)\Big\}, [\Theta_s(x),\Phi_s(y)] = -\ri\theta(x-y)\nonumber\\
&& \Phi_s^{(+)} = \varphi_{1s} +\bar\varphi_{2s}, ~~ \Phi_s^{(-)} = \bar\varphi_{1s} +\varphi_{2s}. \label{H}
\eea
As I have said, at $g_{12} >0$ the cosine operator is relevant and  the spin sector has gap which we denote as  $M_s$. This leads to a sharp increase in the charge and pairing susceptibilities on the corresponding wave vectors. In the formal language this follows from the fact that the spin part of the order parameter operators acquires a finite  vacuum expectation value and hence at energies much smaller than the spin gap one can   rewrite the matrix field (\ref{g}) as 
\bea
\hat g_{(\pm)} = A \left(
\begin{array}{cc} 
\exp[\ri\sqrt{2\pi}\Phi_{(\pm)}] & \exp[\ri\sqrt{2\pi}\Theta_{(\pm)}] \\
-\exp[-\ri\sqrt{2\pi}\Theta_{(\pm)}] & \exp[-\ri\sqrt{2\pi}\Phi_{(\pm)}] 
\end{array}
\right). \label{spin}
\eea
where $\Phi_{(+)},\Theta_{(+)} = \varphi_{c1} \pm \bar\varphi_{c2}, ~~\Phi_{(-)},\Theta_{(-)} = \varphi_{c2}\pm \bar\varphi_{c1}$. In a strictly one-dimensional model such field would not condense, but just display power law correlations at zero temperature and correlation length $\sim 1/T$ at finite T. The scaling dimensions of the CDW and PDW exponents are $d = 1/2 + O(\gamma)$ and for a weak forward scattering are approximately equal. Although Abelian bosonization used to derive this result looks simpler, it hides the rich symmetry of the critical modes present at $\gamma =0$. This symmetry becomes manifest in the non-Abelian bosonization formalism which I will use below to describe coupled chains.

 Simple calculations show that interchain tunneling and the backward scattering due to the Coulomb interaction generate coupling between the order parameters of different chains. It is interesting that the momentum conservation prevents a coupling of OPs with different parity. Depending on what interaction prevails this coupling will lead either incommensurate CDW order or unidirectional PDW one. If the anisotropy between the order parameters  is strong the description of the corresponding Ginzburg-Landau and topological excitations can be found in \cite{review}. The situation with of weak anisotropy when the scaling dimensions of the OPs are approximately the same,  is more interesting. Here we have four intertwined order parameters  - two CDW  ones with identical wave vectors and two PDWs with wave vectors of opposite sign. The strongest interaction fixes the wave vector components parallel to the chains: $Q^z_{cdw} = k_{F1} + k_{F2}$ and $Q^z_{pdw} = k_{F1}-k_{F2}$. The perpendicular components are determined by the interchain interactions. Since the Josephson coupling matrix element is proportional to the product of hopping matrix elements for electrons from the different bands $J_{\perp} \sim t_1t_2/M_s$, one can imagine the situation when $t_1t_2 < 0$ (which, by the way, seems to be the case in the blue bronze) and $J_{\perp}<0$. Then the phase of PDW will oscillate between different chains as it was suggested in \cite{Kivelson} for La$_{1.875}$Ba$_{0.125}$CuO$_4$. 
 
 The effective Ginzburg-Landau  description valid for energies less than the spin gap $M_s$ admits different formulations, one can use the Abelian one as in Eqs. (\ref{H}, \ref{spin}) or resort to the non-Abelian formulation. Although the latter one looks more complicated, it has the advantage  of making manifest  the approximate SU(2)$\otimes$SU(2) symmetry. Here the low energy Hamiltonian for a system of coupled chains is presented as a sum of coupled SU$_1$(2) Wess-Zumino-Novikov-Witten models (see Appendix B for details). For the sake of compactness I provide the expression for the case of equal Fermi velocities:
 \begin{widetext}
 \bea
 && L = L_+ + L_- - \gamma_{12}\sum_n \int \rd x  \mbox{Tr}(\tau^3 g_+^{-1}\p_x g_+)_n\mbox{Tr}(\tau^3 g_-^{-1}\p_x g_-)_n, \\
&& L_{\pm} = \sum_n W[g_n] + \int \rd x \sum_{n,m} \Big\{ J_{\parallel}^{nm}\sum_{a=0,3}\Big[\mbox{Tr}(\tau^a g_n)\mbox{Tr}(\tau^a g^{-1}_m) +h.c.] - J_{\perp}^{nm}\sum_{a=1,2}\Big[\mbox{Tr}(\tau^a g_n)\mbox{Tr}(\tau^a g^{-1}_m) +h.c.] \Big\},\nonumber
\eea
\end{widetext}
where $g_{\pm}(t,x;n)$ are SU(2) matrix fields, $\tau^a$ are Pauli matrices and $W[g]$ is the SU$_1$(2) WZNW Lagrangian described in the Appendix. 

  At last I would like to comment on the methods of detection. The most popular method to detect PDW is scanning tunneling microscopy with normal (STM) and superconducting tip (Josephson STM or JSTM). STM measures the local density of states and it is suggested that PDW will produce periodic changes in the coherence peaks. This will not occur in the present model. Indeed,  the spectral gap in the present model is the spin gap, it is a high energy feature which is not influenced  by CDW or PDW. The spatial variations of the local DOS is proportional to the CDW order parameter.  The only suitable method is JSTM which measures the square of the critical current $I_j^2(r)$ and the current does oscillate in space: $I_j(r) \sim \Delta_{pdw}^{(+)}\re^{\ri Qr} + \Delta_{pdw}^{(-)}\re^{-iQr}$.
  
  {\it Conclusions. }  I have described a simple weak  coupling mechanism of PDW formation. Although this mechanism can work in any space dimension, the possible experimental realization I am aware about is quasi one dimensional.  I am presented a simple analytically tractable model describing PDW formation and identified a class of materials which it may describe.  In the suggested context  a unidirectional PDW is intertwined with CDW. Their wave vectors are not related to each other. 
    
 {\it Acknowledgements}

 I am grateful to A. Chubukov and I. Zaliznyak for valuable remarks. The work  was supported by Office of Basic Energy Sciences, Material Sciences and Engineering Division, U.S. Department of Energy (DOE) under Contract No. DE-SC0012704.

\begin{widetext}
\section{Appendix}

\subsection{Order parameters. Abelian bosonization} 

The bosonization rules for the fermion operators are 
\bea
R_{a,\s} = \frac{\xi_{a,\s}}{\sqrt{2\pi a_0}}\re^{-\ri\sqrt{2\pi}(\varphi_{a,c} + \s\varphi_{a,s})}, ~~ L_{a,\s} = \frac{\xi_{a,\s}}{\sqrt{2\pi a_0}}\re^{\ri\sqrt{2\pi}(\bar\varphi_{a,c} + \s\bar\varphi_{a,s})}, \label{fermions}
\eea
where $\xi_{a,\s}$ are Klein factors $\{\xi_{a,\s},\xi_{b,\s'}\} = 2\delta_{\s\s'}\delta_{a,b}$ and $\varphi_{a,s}, ~~\varphi_{a,c}, ~~ \bar\varphi_{a,s}, ~~\bar\varphi_{a,c}$ are chiral charge and spin fields governed by the Gaussian action. We can represent them via Pauli matrices: 
\bea
\xi_{1\uparrow} = \s^x\otimes\tau^z, ~~ \xi_{1\downarrow} = \s^y\otimes\tau^z, ~~ \xi_{2\uparrow} = (I\otimes\tau^x), ~~ \xi_{2\downarrow}= (I\otimes\tau^y),
\eea
and fix the irreducible representation of the Clifford algebra by imposing   condition $\s^z\otimes\tau^z = 1$.

The order parameters for the $+$ sector are
\bea
&& \Delta_{pdw}(+) = R_{1\uparrow}L_{2\downarrow} - R_{1\downarrow}L_{2\uparrow} = \frac{1}{2\pi a_0}\re^{\ri \sqrt{2\pi}(-\varphi_{1c}+\bar\varphi_{2c})}\Big[ \xi_{1,\uparrow}\xi_{2\downarrow}\re^{-\ri\sqrt{2\pi}(\varphi_{1s} + \bar\varphi_{2s})} - \xi_{1,\downarrow}\xi_{2\uparrow}\re^{\ri\sqrt{2\pi}(\varphi_{1s} + \bar\varphi_{2s})}\Big],\label{pdw} \\
&& \Delta_{cdw}(+) = R^+_{1\s}L_{2\s} = \frac{1}{2\pi a_0} \re^{\ri\sqrt{2\pi}(\varphi_{1c}+\bar\varphi_{2c})}\Big[\xi_{1\uparrow}\xi_{2\uparrow}\re^{\ri\sqrt{2\pi}(\varphi_{1s}+\bar\varphi_{2s})} + \xi_{1\downarrow}\xi_{2\downarrow}\re^{-\ri\sqrt{2\pi}(\varphi_{1s}+\bar\varphi_{2s})}\Big]. \label{cdw}
\eea
Imposing the gauge fixing projector $P = (1+\s^z\otimes\tau^z)/2$ we get 
\bea
&& \Delta_{pdw}(+) = \frac{1}{2\pi a_0}\re^{\ri \sqrt{2\pi}(-\varphi_{1c}+\bar\varphi_{2c})}\sin[\sqrt{2\pi}(\varphi_{1s}+\bar\varphi_{2s})](-\s^x\otimes\tau^x + \s^y\otimes\tau^y), \\
&&  \Delta_{cdw}(+) = \frac{1}{2\pi a_0}\re^{\ri \sqrt{2\pi}(\varphi_{1c}+\bar\varphi_{2c})}\sin[\sqrt{2\pi}(\varphi_{1s}+\bar\varphi_{2s})](\s^x\otimes\tau^y +\s^y\otimes\tau^x).
\eea
The interaction is 
\bea
V_{int} = g_{12}{\bf J}_{R1}{\bf J}_{L2} = \frac{g}{\pi}\p\varphi_{1s}\bar\p\bar\varphi_{2s} + \frac{2g}{(2\pi a_0)^2} (\s^z\otimes\tau^z)\cos[\sqrt{8\pi}(\varphi_{1s}+\bar\varphi_{2s})],
\eea
where $\p = \frac{1}{2}(\p_{\tau} - \ri\p_x), ~~ \bar\p = \frac{1}{2}(\p_{\tau} + \ri\p_x)$ and $v=1$ for simplicity. The vacuum in the sector $\s^z\otimes\tau^z =1$ is $\Phi^{(+)}_s = \sqrt{8\pi}(\varphi_{1s} + \bar\varphi_{2s}) = \pi$ so that $\la \sin(\Phi/^{(+)}2) \ra \neq 0$ and the amplitudes of the order parameters (\ref{pdw},\ref{cdw}) are finite.

At nonzero $\gamma$ assuming that $K_{11} = K_{22} = K$ and $\gamma_{12} =0$, we have 
\bea
\Delta^{(\pm)}_{pdw,cdw} \sim \exp\{\ri\sqrt{\pi/2}[(\sqrt K + 1/\sqrt K)(\varphi_{1c} \pm \bar\varphi_{2c}) + (\sqrt K-1\sqrt K)(\pm \varphi_{1c} +\bar\varphi_{2c})]\}.
\eea
CDW and PDW have the same scaling dimensions. 

\subsection{Non-Abelian formulation} 

The non-Abelian formulations makes the SU$_R$(2)$\otimes$SU$_L$(2) symmetry of the charge sector manifest. 

At $K=1$ the charge sectors $(\pm)$ possess SU(2) symmetry. For the $(+)$ sector it is  generated by the current operators 
\bea
&& J^+ = R^+_{\uparrow,1}R^+_{\downarrow,1} = \frac{\ri}{2\pi a_0}\re^{-\ri\sqrt{8\pi}\varphi_{c1}}, ~~ \bar J^+ = L^+_{\uparrow,1}L^+_{\downarrow,2} = \frac{\ri}{2\pi a_0}\re^{\ri\sqrt{8\pi}\varphi_{c2}},\\
&& J^- = R^+_{\downarrow,1}R^+_{\uparrow,1}= -\frac{\ri}{2\pi a_0}\re^{\ri\sqrt{8\pi}\varphi_{c1}}, ~~ \bar J^- = L^+_{\downarrow,1}L^+_{\uparrow,2} = -\frac{\ri}{2\pi a_0}\re^{-\ri\sqrt{8\pi}\varphi_{c2}},\\
&& J^3 = :R^+_{\s,1}R_{\s,1} = \frac{\ri}{\sqrt{2\pi}}\p\varphi_{c1}, ~~ \bar J^3 = :L^+_{\s,2}L_{\s,2}: = -\frac{\ri}{\sqrt{2\pi}}\bar\p\bar\varphi_{c2}.
\eea
These operators satisfy SU$_1$(2) Kac-Moody algebra and the entire action can be recast in terms of the Wess-Zumino-Novikov-Witten model. In this formulation the dynamics of the matrix field $g$ (\ref{spin}) is governed by the Lagrangian (to avoid cumbersome expressions I set both Fermi velocities $v$ equal to 1).

\bea
W[g] = \frac{1}{16\pi}\int  \rd x \mbox{Tr}(\p_{\mu}g^+\p_{\mu} g) - \frac{\ri}{24\pi}\int_0^{\infty}\rd\xi \int \rd x\epsilon^{\alpha\beta\gamma}\mbox{Tr}(g^+\p_\alpha gg^+\p_\beta gg^+\p_\gamma g).
\eea


\end{widetext}
\end{document}